\documentclass[useAMS,usenatbib]{mn2e}
\usepackage{graphicx}
\usepackage{txfonts}
\usepackage{natbib}

\newcommand{\kms}    {km~s$^{-1}$}
\newcommand{\jyb}    {Jy~beam$^{-1}$}
\def\araa{ARA\&A}%
\def\apj{ApJ}%
\def\apjl{ApJ}%
\def\apss{Ap\&SS}%
\def\aap{A\&A}%
\def\mnras{MNRAS}%
\def\pasj{PASJ}%
\def\nat{Nature}%
\title[Magnetic field regulated protostellar infall]{Magnetic field regulated infall on the disc around the massive protostar Cepheus~A~HW2}
\author[W. H. T. Vlemmings et al.]{W. H. T. Vlemmings$^{1}$\thanks{E-mail: wouter@astro.uni-bonn.de}, G. Surcis$^{1,2}$, K. J. E. Torstensson$^{3,4}$ and H. J. van Langevelde$^{4,3}$\\
$^{1}$Argelander-Institut f{\"u}r Astronomie, University of Bonn, Auf dem H{\"u}gel 71, D-53121 Bonn, Germany\\
$^{2}$Max-Planck Institut f{\"u}r Radioastronomie, Auf dem H{\"u}gel 69, D-53121 Bonn, Germany\\
$^{3}$Leiden Observatory, Leiden University, PO Box 9513, NL-2300 RA Leiden, The Netherlands\\
$^{4}$Joint Institute for VLBI in Europe, PO Box 2, NL-7990 AA Dwingeloo, The Netherlands}

\begin{document}

\date{05-01-2010}

\pagerange{\pageref{firstpage}--\pageref{lastpage}} \pubyear{2009}

\maketitle

\label{firstpage}

\begin{abstract}

  We present polarization observations of the 6.7-GHz methanol masers
  around the massive protostar Cepheus~A~HW2 and its associated
  disc. The data were taken with the Multi-Element Radio Linked
  Interferometer Network. The maser polarization is used to determine
  the full three-dimensional magnetic field structure around
  Cepheus~A~HW2. The observations suggest that the masers probe
  the large scale magnetic field and not isolated pockets of a
  compressed field. We find that the magnetic field is predominantly
  aligned along the protostellar outflow and perpendicular to the
  molecular and dust disc. From the three-dimensional magnetic field
  orientation and measurements of the magnetic field strength along
  the line of sight, we are able to determine that the high density
  material, in which the masers occurs, is threaded by a large scale
  magnetic field of $\sim23$~mG. This indicates that the protostellar
  environment at $\sim1000$~AU from Cepheus~A~HW2 is slightly
  supercritical ($\lambda\approx1.7$) and the relation between density
  and magnetic field is consistent with collapse along the magnetic
  field lines. Thus, the observations indicate that the magnetic field
  likely regulates accretion onto the disc. The magnetic field
  dominates the turbulent energies by approximately a factor of three
  and is sufficiently strong to be the crucial component stabilizing
  the massive accretion disc and sustaining the high accretion rates
  needed during massive star-formation.
\end{abstract}


\section{Introduction}

Massive stars are short lived, yet they dominate galaxy evolution due
to their strong radiation while enriching the interstellar medium with
heavy elements when they explode in a supernova. Massive stars
typically form in distant, dense clusters. In such regions,
gravitational, radiation and turbulent energies are different from
those in the regions that form lower mass stars such as the
Sun. Specifically, young stars more massive than 8 solar masses
($M_\odot$) have a radiation pressure that would be sufficient to halt
infall and prevent the accretion of additional mass
\citep[e.g.][]{Wolfire87}. Still, stars more massive than 200
$M_\odot$ have been observed \citep[e.g.][]{Figer98}, indicating that
the processes governing massive star-formation are yet poorly
understood.

A number of different formation scenarios have been proposed
\citep{Zinnecker07}. These include formation through the merger of
less massive stars \citep{Bonnell05} or through the accretion of
unbound gas from the molecular cloud, the competitive accretion theory
\citep{Bonnell98}. In the third scenario, core accretion
\citep{McKee03}, massive stars form through gravitational collapse,
which involves disc-assisted accretion to overcome radiation
pressure. This scenario is similar to the favored picture of low-mass
star-formation \citep{Shu95}, in which magnetic fields are thought to
play an important role by removing excess angular momentum, thereby
allowing accretion to continue onto the star \citep{Shu95,
  Basu94}. However, the accretion rates during massive
star-formation are significantly higher than during the formation of a
low-mass star, and it is unclear whether massive accretion discs are
able to sustain these rates without support by strong magnetic fields
\citep{McKee03}. Still, if such magnetized discs exist around
massive protostars, outflows will arise as a natural consequence
\citep{Banerjee07}.

Current observations of magnetic fields in massive star formation
regions are often limited to low density regions (hydrogen number
density $n_{\rm H_2}<10^6$cm$^{-3}$) and/or envelopes at scales of
several 1000~AU. Linear polarization observations of dust also only
provide information on the magnetic field in the plane of the sky, so
they have been yet unable to probe the strength and the full structure
of the magnetic field close to the protostar and around protostellar
discs. Recent high-angular-resolution submillimeter dust polarization
observations of the massive hot molecular core G31.41+0.31 indicate
that the gravitational collapse shows signs of being controlled by the
magnetic field scales of $\sim 5000$~AU \citep{Girart09}. Currently, the only
probes of magnetic fields in the high density regions close to the
massive protostars are masers. Because of their compactness and high
brightness, masers are specifically suited for high angular resolution
observations of weak linear and circular polarization and can thus be
used to determine the magnetic field strength and morphology down to
AU scales \citep[see][for a review of maser
polarization]{Vlemmings07a}. Most observations have focused on OH and
H$_2$O masers \citep[e.g.][]{Bartkiewicz05, Vlemmings06}, but
recently, methanol masers have been shown to be excellent probes of
the magnetic field \citep[e.g.][]{Vlemmings06c, Vlemmings08a,
  Surcis09}. A number of both OH and methanol maser observations
indicate that, although individual maser features are only a few AU in
size, they probe the large scale magnetic field around the massive protostar \citep[e.g.][]{Surcis09}.

We observed Cepheus A, which, at a distance of 700 pc
\citep{Moscadelli09}, is one of the closest regions of active massive
star-formation. Located in the Cepheus OB3 complex, it hosts a
powerful extended bipolar molecular outflow that likely originates
from HW2, the brightest radio continuum source in the region
\citep{Hughes84}. HW2 is thought to be a young protostar of spectral
type B0.5 with a mass of $\sim$20 $M_\odot$ \citep{Jimenez07}. The
extended outflow appears to be driven by a small scale ($\sim$700 AU)
thermal radio jet, with ionized gas ejected at a velocity of
$\sim$500~\kms \citep{Curiel06}. In addition to the bipolar outflow,
Cepheus A HW2 is surrounded by a rotating disc of dust and molecular
gas oriented perpendicular to the jet \citep{Jimenez07, Patel05,
  Torrelles07}. The jet and disc structure supports the picture where
massive stars form through disc accretion, similar to low-mass
stars. The environment of Cepheus A HW2 however, is significantly more
complex than that of typical low-mass protostars. Besides HW2, at
least three additional young stellar objects are located within an
area of $\sim$$500\times500$ AU. All these sources seem to be located
within or close to the molecular disc \citep{Torrelles07, Comito07,
  Jimenez09}. The presence of OH, H$_2$O and CH$_3$OH (methanol)
masers in or near to the disc of Cepheus~A~HW2 gives us an unique
opportunity to study its magnetic field \citep{Vlemmings06,
  Bartkiewicz05}, with recent methanol maser Zeeman-splitting
  measurements revealing a dynamically important magnetic field with a
  $\sim8$~mG component along the maser
  line-of-sight \citep{Vlemmings08a}.

In \S~\ref{obs} we present the methanol maser observations and in
\S~\ref{method} we shortly discuss the radiative transfer tool used to
infer the full 3-dimensional magnetic field geometry. \S~\ref{results}
presents the results of our observations and in \S~\ref{disc} we
discuss the derived 3-dimensional magnetic field morphology and what
its implication for the role of the magnetic field during high-mass
star-formation. Additionally, we highlight the potential for methanol
maser polarization observations of the upgraded Multi-Element Radio
Linked Interferometer Network (e-MERLIN).

\begin{figure}
\includegraphics[width=8.0cm]{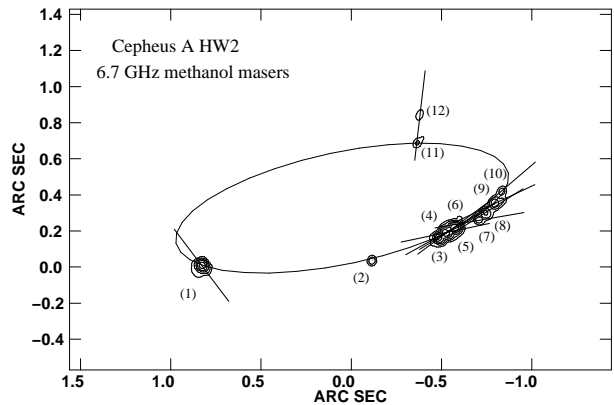}
\caption[spectra]{Total intensity (I) map of the 6.7-GHz methanol masers around Cepheus~A~HW2. Contours are drawn at $5, 10, 20, 40~\&~80$~\jyb. The labels identify the features presented in Tables~\ref{tab1} and \ref{tab2}. The vectors indicate the observed polarization angle $\chi$. Also indicated is the best fit ellipse to the methanol maser feature distribution observed with MERLIN.}
\label{Fig:cepa}
\end{figure}

\section{Observations}

\label{obs}

The $5_1-6_0$~A$^+$ maser transition of methanol at 6.7-GHz was
observed on 2006 December 2-4 using 6 of the MERLIN telescopes. Total
on source time was $\sim27$~hr, which was interleaved with
observations of the phase-calibrator source 2300+638. The resulting
beam-size was $40\times30$~mas. We used a 250~kHz bandwidth with 256
channels for a total velocity coverage of $\sim11$~\kms~ centered on
the source velocity $V_{\rm LSR}=-3$~\kms. This provided a velocity
resolution of $0.044$~\kms. To obtain the linear polarization, the
data were correlated with the full Stokes parameters. For calibration
purposes, the continuum calibrators were observed with the 16~MHz
wide-band mode. Both 3C84 and 3C286 were also observed in the
narrow-band spectral line configuration and were used to determine the
phase off-set between the wide- and narrow-band setup. Flux and
bandpass calibration was done using the primary flux calibrator
3C286. Instrumental feed polarization was determined using the
unpolarized calibrator 3C84 and the polarization angle was calibrated
using 3C286. The residual instrumental linear polarization was
determined to be less than $<0.1$~percent from the observations of
3C84 and 2300+638. Self-calibration, in RCP and LCP separately, was
performed on the strongest isolated maser feature. After calibration,
the antenna contributions were re-weighted according to their
sensitivity at 5~GHz and their individual efficiency. Finally,
$5.12\times5.12"$ image cubes were created in Stokes I, Q, and U. The
rms noise in the emission free channels was $\sim6$~m\jyb~for Stokes I
and $\sim9$~m\jyb~for Stokes Q and U. For the channels with strong
maser emission we are limited to a dynamic range of $\sim850$ due to
residual calibration errors. Because of the dynamic range limit we
were unable to determine the circular polarization due to Zeeman
splitting in these observations. The non-detection of the
$\sim0.17$~percent circular polarization observed for the 6.7-GHz
methanol masers using the Effelsberg 100-m telescope
\citep{Vlemmings08a} is unsurprising considering the rms noise limits
of our circular polarization observations ($1\sigma\sim0.12$ percent).

\section{Maser polarization radiative transfer}
\label{method}

Maser theory describes that the fractional linear polarization $P_l$
of the non-paramagnetic maser species such as H$_2$O, SiO and
CH$_3$OH, depends on the degree of maser saturation and the angle
$\theta$ between the maser propagation direction and the magnetic
field. Additionally, the relation between the measured polarization
angle $\chi$ and the magnetic field angle on the sky $\phi_B$ also
depends on $\theta$, with the polarization vectors generally parallel
to the field for $\theta<\theta_{\rm crit}\approx55^\circ$ and
perpendicular to the field for $\theta>\theta_{\rm crit}$
\citep{Goldreich73}. However, non-magnetic effects, such as full maser
saturation and anisotropic maser pumping, can generate enhanced linear
polarization and can rotate the polarization vectors with respect to
the magnetic field \citep{Deguchi90}. Thus, before we can use
polarization observations to determine the magnetic field morphology,
we need to estimate the level at which these effects influence our
measurements. The rotation and generation of linear polarization
depends on the amount of Zeeman-splitting and the level of maser
saturation described by the Zeeman frequency shift $g\Omega$, the
maser decay rate $\Gamma$ and the stimulated emission rate $R$. When
$g\Omega>>\Gamma>R$, the linear polarization direction is determined
by the magnetic field, even in the presence of anisotropic
pumping. When the effect of anisotropic pumping is not significant, even $g\Omega>\Gamma$ is a sufficient condition \citep{Watson02}. The maser stimulated emission rate is given by:
\begin{equation}
R \simeq A k T_{\rm b}\Delta\Omega / 4\pi h\nu.
\end{equation}
Here $A$ is the Einstein coefficient for the maser transition, which
is equal to $0.1532\times10^{-8}$~s$^{-1}$, and $k$ and $h$ are the
Boltzmann and Planck constants respectively. The maser frequency is
denoted by $\nu$, and $T_{\rm b}$ and $\Delta\Omega$ are the maser
brightness temperature and beaming solid angle respectively. We
measure $T_b<10^{10}$~K for our masers assuming that they are almost
resolved with the MERLIN beam of $\sim30$~mas. However, when observed
at even higher angular resolution, more typical sizes are of the order
of a few mas and brightness temperatures reach of order $10^{12}$~K
\citep{Menten92}. The beaming angle $\Delta\Omega$
cannot easily be measured directly. Still, those of H$_2$O masers in star-forming
regions are shown to be $\Delta\Omega\sim10^{-4}-10^{-5}$
\citep{Nedoluha91}, while those in the envelopes of evolved stars are
of order $10^{-2}$ \citep{Vlemmings05}. It is likely that methanol
masers have similar beaming angles. Taking $\Delta\Omega\sim10^{-2}$
as a conservative upper limit, the maser stimulated emission rate
$R<0.04$~s$^{-1}$.

The maser decay rate $\Gamma$ is more difficult to determine. For
maser transitions involved infrared pumping, which includes methanol,
$\Gamma\sim1$~s$^{-1}$ \citep{Scap92}. However, infrared trapping can
reduce this value, although likely not by more than an order of
magnitude \citep{Goldreich74}. Alternatively, when the maser
transition involves collisions, $\Gamma=0.1 n_{H_2}/10^9$~s$^{-1}$,
with $n_{H_2}$ the hydrogen number density in cm$^{-3}$. For the
methanol masers this would imply $\Gamma<0.1$~s$^{-1}$. However, as
the collisional decay rate is smaller than the radiative decay rate,
the radiative decay rate determined $\Gamma$. We thus assume
$\Gamma=1$~s$^{-1}$, though other authors have assumed the lower value
of $\Gamma=0.6$~s$^{-1}$ \citep{Minier02}.

We have thus shown that even with conservative estimates for $R$ and
$\Gamma$, we are in the regime where $\Gamma>R$. However, we need to
asses if $g\Omega$ is sufficiently much larger than $R$ and $\Gamma$
so that, even in the case of anisotropic pumping, the polarization
vectors can be used to determine the magnetic field direction. The
analysis by \citet{Nedoluha90} indicates that the effect of
anisotropic pumping decreases significantly for transitions with high
angular momentum, such as the 6.7-GHz methanol masers. This is further
confirmed by the fact that we do not observe any relation between the
maser total intensity and fractional polarization, which would be
expected when non-magnetic effects contribute to the
polarization. However, even without having to consider anisotropic
pumping, the quantization axis of the maser is potentially not
sufficiently determined by the magnetic field when $g\Omega$ is not
significantly larger than $R$ and $\Gamma$.  For a magnetic field
strength of $>8$~mG in the Cepheus~A methanol maser region
\citep{Vlemmings08a}, $g\Omega>10$~s$^{-1}$. Although the uncertainty
on the magnetic field determination was estimated to be of order
$25$~percent, the limit of $8$~mG assumes a magnetic field along the
maser line of sight and the true magnetic field, and consequently
$g\Omega$, will thus likely be larger.  \citet{Nedoluha90} found that,
for a ratio $R/\Gamma<0.1$ and $g\Omega/\Gamma>10$, the relation
between polarization angle and magnetic field direction hold to within
$\sim15^\circ$. This improves quickly to better than $\sim5^\circ$
when $g\Omega/\Gamma$ becomes larger. We are thus in the regime where
the polarization vectors directly determine the magnetic field
direction, but adopt an additional $5^\circ$ uncertainty in our
determinations of the magnetic field direction on the plane of the
sky.

For H$_2$O masers it has been shown to be possible to estimate the
saturation level, as determined by the emerging brightness temperature
$T_b\Delta\Omega$, and the intrinsic maser line-width $\Delta V_i$ by
fitting both the total intensity and linear polarization maser
spectra \citep{Vlemmings06}. As $T_b\Delta\Omega$ determines the relation between $P_l$
and $\theta$ we are then able to constrain $\theta$ using the measured
$P_l$ and thus provide the full 3-dimensional magnetic field
direction. We have adapted the same modeling tools used for the 22~GHz
H$_2$O masers to model the 6.7-GHz methanol maser transition. 

\begin{figure*}
\includegraphics[width=16.0cm]{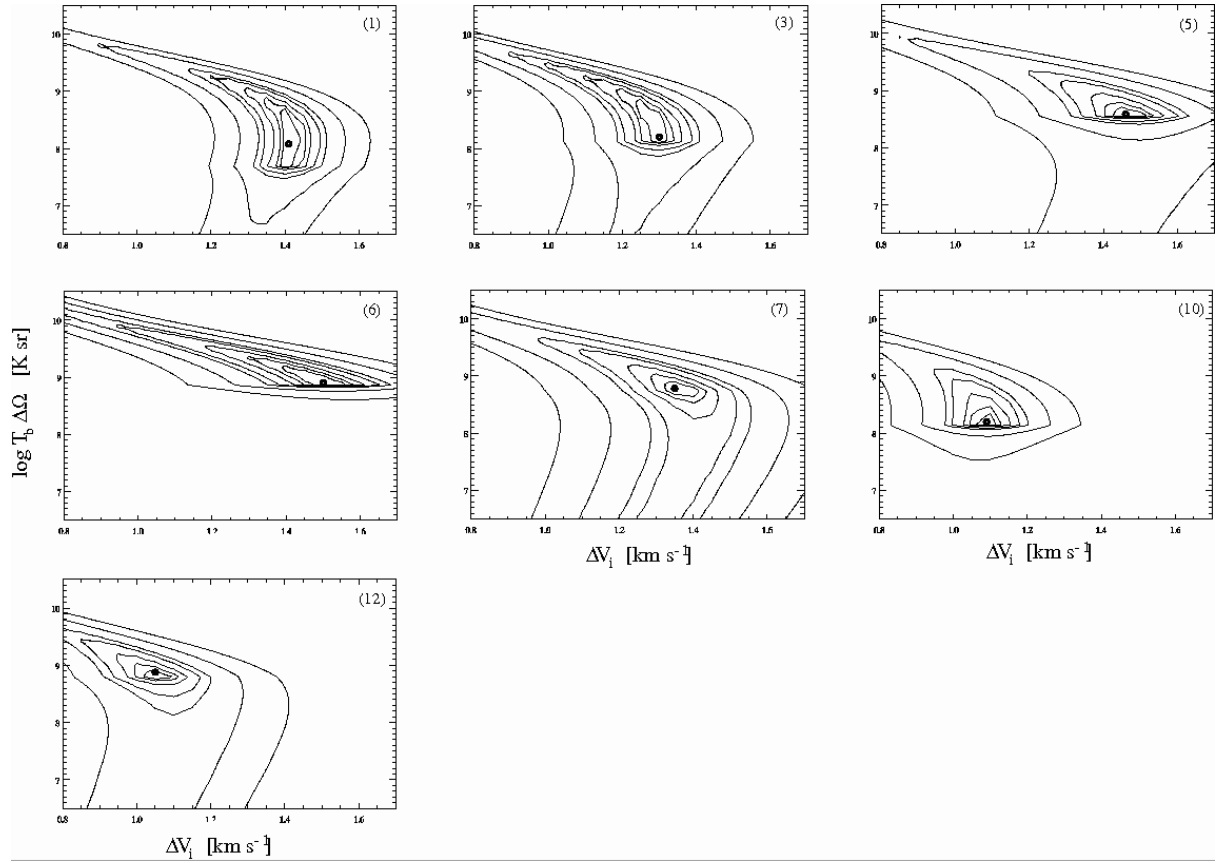}
\caption[modelfit]{Results of the full radiative transfer
  $\chi^2$-model fits, for those maser features from
  Fig.~\ref{Fig:spectra} that do not suffer from blending and for
  which linear polarization was detected. The fits yield the emerging
  maser brightness temperature $T_b\Delta\Omega$ and intrinsic maser
  line-width $\Delta V_i$. Contours indicate the significance
  intervals $\Delta\chi^2=0.25,0.5,1,2,3,7$, with the thick solid
  contour indicating the $1\sigma$ area.}
\label{Fig:fit}
\end{figure*}

\begin{figure*}
\includegraphics[width=16.0cm]{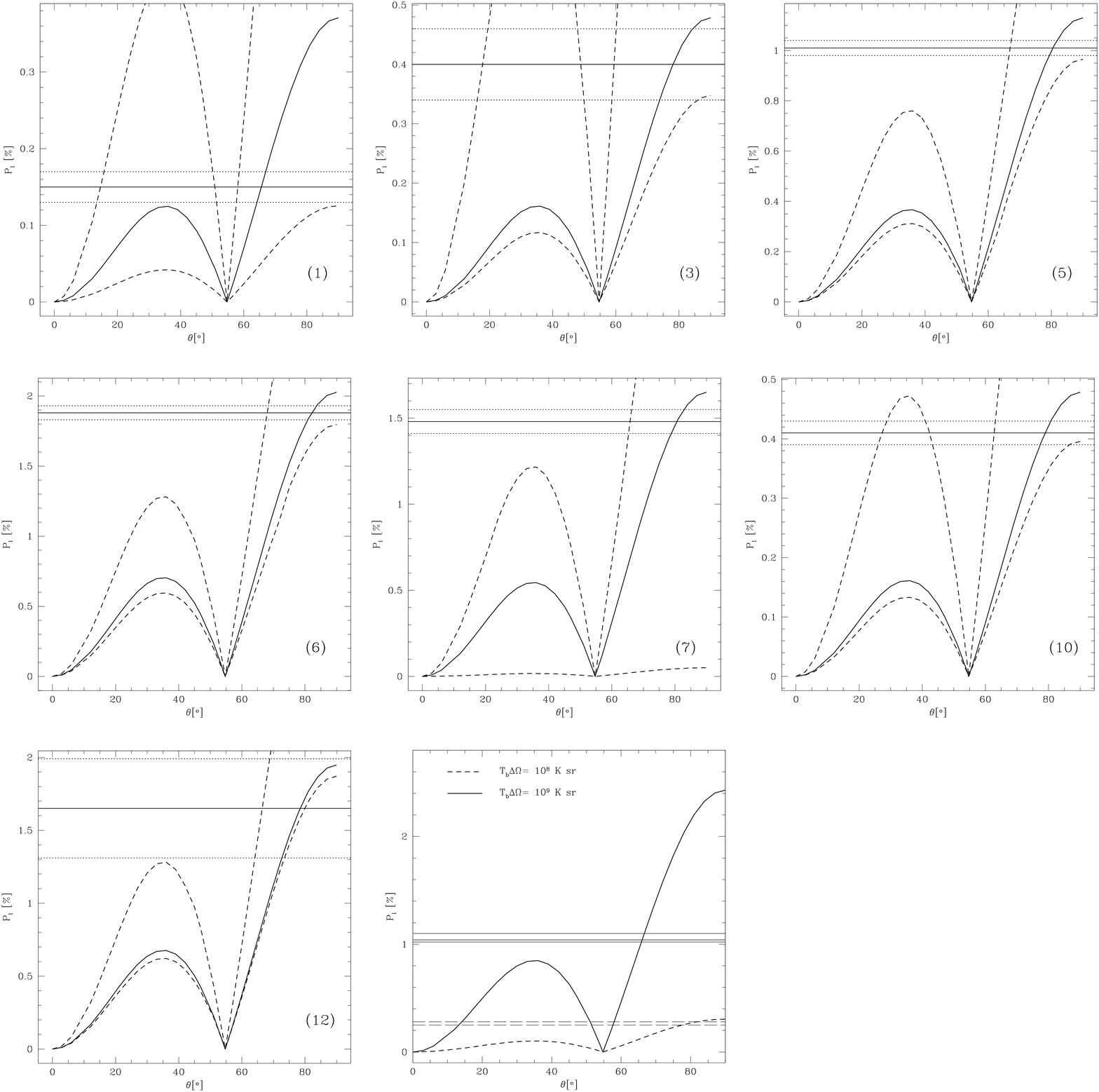}
\caption[thetafit]{The angle $\theta$ between the maser propagation direction and the magnetic field vs. the fractional linear polarization $P_l$. Panels are presented for the $7$ maser features for which an accurate fit could be made to the line profiles. The solid curves indicate the best fit emerging brightness temperatures and the dashed curves are the $\pm1\sigma$ intervals. The horizontal solid and dashed lines are the measured polarization fraction and the corresponding errors. The final panel indicates the remaining $5$ maser features with curves for typical emerging brightness temperatures. In this panel, the dashed horizontal lines are the $3\sigma$ upper limits for the maser features without measured polarization, as these limits provide strong constraints on $\theta$.}
\label{Fig:thetafit}
\end{figure*}

\section{Results}
\label{results}

\subsection{Maser distribution}
\label{dist}

\begin{figure*}
\includegraphics[width=16.0cm]{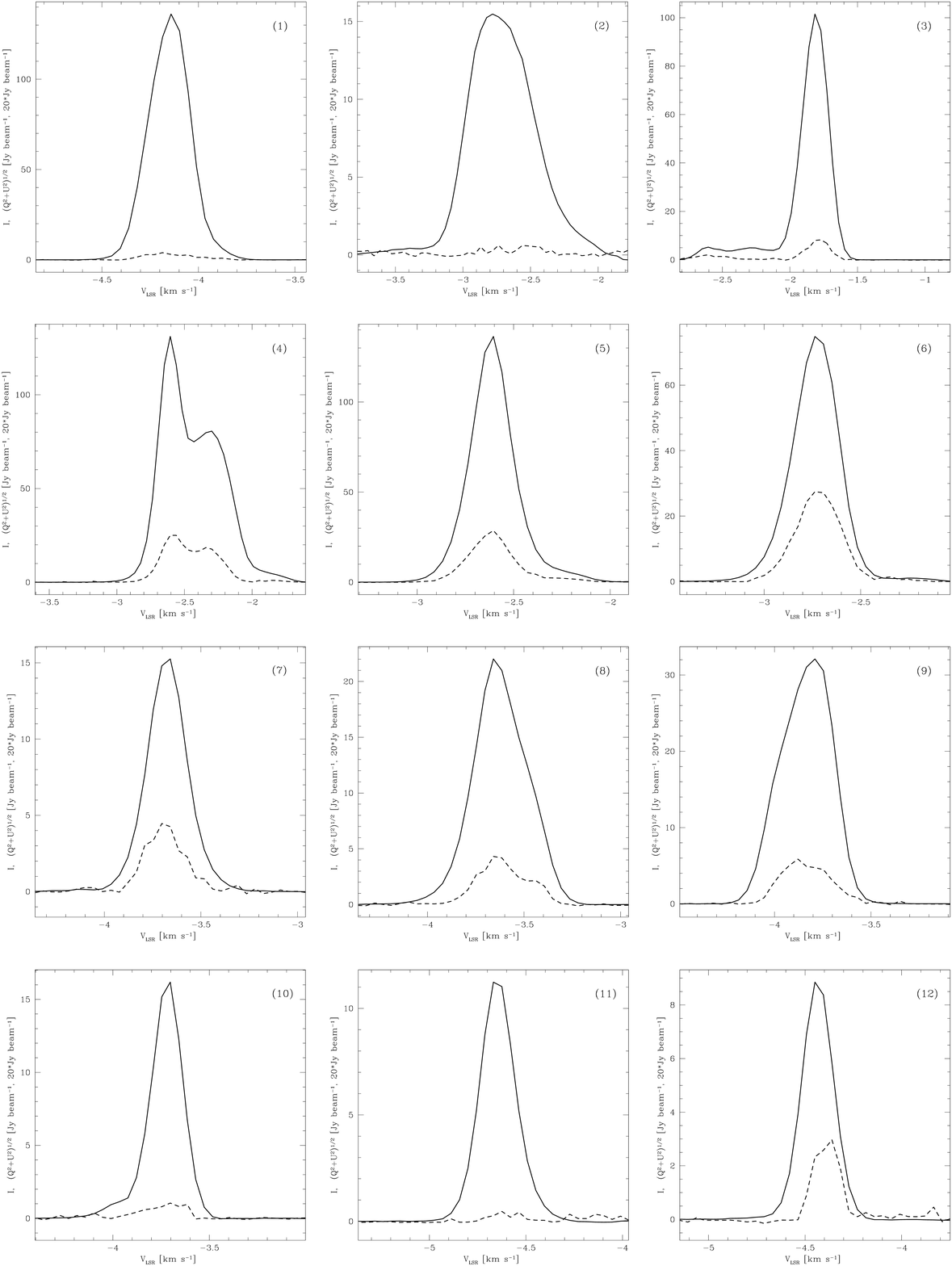}
\caption[spectra]{Total intensity (I; solid) and linear polarization (dashed) spectra for the 12 maser features detected above a level of $5$~\jyb~in the MERLIN observations. The linear polarization spectra have been multiplied by a factor $20$.}
\label{Fig:spectra}
\end{figure*}

The resulting velocity integrated image of the
6.7-GHz methanol masers around Cepheus~A~HW2 is shown in
Fig.~\ref{Fig:cepa}. No maser
emission was detected at $<5$~\jyb. The $12$ maser features that were
identified are shown in Fig.~\ref{Fig:spectra} and presented in
Table~\ref{tab1} with position offsets ($\Delta\alpha$ and
$\Delta\delta$), velocity ($V_{\rm LSR}$), peak flux ($I_{\rm peak}$),
fractional linear polarization ($P_l$) and polarization angle
($\chi$).

The 6.7-GHz methanol masers make up a nearly perfect elliptical
distribution around HW2 and its disc, and the maser distribution
observed with MERLIN matches that seen in European VLBI Network (EVN)
and Japanese VLBI Network (JVN) observations \citep[Torstensson et al,
in prep.;][]{Sugiyama08}. A least square fit of an ellipse to the
MERLIN data indicates that the masers occur in a ring with a radius of
$\sim$650 AU, a thickness of $\sim$300 AU, a projection angle on the
sky of $\phi=-78^\circ$ and an inclination angle of $i=71^\circ$. The
maser ring appears tilted with respect to the molecular ($R=580$~AU,
$\phi=-56^\circ$ and $i=68^\circ$) and dust disc $(R=330$~AU,
$\phi=-59^\circ$ and $i=56^\circ$). Observations at high resolution
($9\times7$ mas) with the EVN further indicate that the maser
velocities are best fit by infall towards HW2 at $\sim$1.7~\kms
\citep[][Torstensson et al., in prep.]{Kalle08}. While the molecular
disc rotates with a velocity of $\sim$5~\kms \citep{Jimenez07}, no
significant rotation is observed in the maser region.

\begin{table*}
\caption{Observation Results}
\centering
\begin{tabular}{lcccccc}
\hline\hline
Feature & $\Delta\alpha$ & $\Delta\delta$ & $V_{\rm LSR}$ & $I_{\rm peak}$ & $P_l$ & $\chi$ \\
 & (arcsec)$^a$ & (arcsec)$^a$ & (\kms) & (\jyb) & (percent) & ($^\circ$) \\
\hline
1 & 0.826 & 0.010 & -4.14 & 136.1 & $0.15\pm0.02$ & $37\pm6$ \\
2 & -0.113 & 0.038 & -2.78 & 15.5 & $<0.75$ & $-$ \\
3 & -0.489 & 0.161 & -1.82 & 101.5 & $0.40\pm0.06$ & $-79\pm2$ \\
4 & -0.528 & 0.189 & -2.30 & 131.1 & $1.10\pm0.08$ & $-65\pm10$ \\
5 & -0.564 & 0.212 & -2.60 & 136.3 & $1.01\pm0.03$ & $-56\pm2$ \\
6 & -0.593 & 0.228 & -2.73 & 74.8 & $1.88\pm0.05$ & $-56\pm4$ \\
7 & -0.710 & 0.270 & -3.68 & 15.3 & $1.48\pm0.07$ & $-80\pm2$ \\
8 & -0.739 & 0.308 & -3.65 & 22.0 & $1.02\pm0.07$ & $-61\pm3$ \\
9 & -0.805 & 0.356 & -3.79 & 32.1 & $1.04\pm0.15$ & $-65\pm1$ \\
10 & -0.830 & 0.430 & -3.70 & 16.2 & $0.41\pm0.02$ & $-49\pm2$ \\
11 & -0.364 & 0.687 & -4.66 & 11.2 & $<0.84$ & $-$ \\
12 & -0.379 & 0.845 & -4.55 & 8.8 & $1.65\pm0.34$ & $-3\pm4$ \\
\hline
\multicolumn{7}{l}{$^a$ Position offsets with respect to $\alpha=22^h56^m17.98^s$, $\delta=62^\circ01'49.390"$}
\end{tabular}
\label{tab1}
\end{table*}

\subsection{Maser polarization and turbulent line-width}
\label{models}

We used the radiative transfer method described in \S~\ref{method} to
determine the emerging brightness temperature, $T_b\Delta\Omega$, and
intrinsic line-width, $\Delta V_i$, of the observed maser features.
As a direct fit requires a regular maser spectrum that does not suffer
from blending, we were able to fit to $7$ of the masers around
Cepheus~A~HW2. The results of the fits are shown in
Figs.~\ref{Fig:fit} and \ref{Fig:thetafit} and given in
Table~\ref{tab2}. They indicate an emerging brightness temperature in
the range $10^8<T_b\Delta\Omega<10^9$~K~sr, implying that the masers
are not saturated, as the ratio of stimulated emission rate over decay
rate $0.04<R/\Gamma<0.4$, and that our assumption for $\Delta\Omega$
in \S~\ref{method} is valid. It is worthwhile to note that our model
calculations of the $T_b\Delta\Omega$ assume a maser decay rate of
$\Gamma\sim1$~s$^{-1}$. However, the model emerging brightness
temperature scale linearly with $\Gamma$ and consequently the
saturation level ($R/\Gamma$) does not change. The intrinsic maser line-width lies
in the range $1.0<\Delta V_i<1.5$~\kms, with a weighted average
$\Delta V_i=1.3\pm0.2$~km/s, which is much larger than the observed
line-width of $<0.3$~\kms~ and shows that re-broadening, as a result
of maser saturation, has not occurred yet. As the temperature needed
to excite the 6.7-GHz masers is $\sim200$~K \citep{Cragg05}, the
thermal line-width is $\sim0.7$~\kms. The intrinsic line-width is thus
significantly broader due to turbulence in the maser region and we
estimate the turbulent line-width to be $\Delta V\approx1.1$~\kms.

\begin{table}
\caption{Fit Results}
\centering
\begin{tabular}{lcccc}
\hline\hline
Feature& $\Delta V_i$ & $T_b\Delta\Omega$ & $\theta$ & $\phi_B$ \\
 & (\kms) & ($\log{K~sr}$) & ($^\circ$) & ($^\circ$) \\
\hline
1 & $1.41^{+0.04}_{-0.08}$ & $8.08^{+0.58}_{-0.47}$ & $47^{+27}_{-24}$ & $37\pm11$\\
3 & $1.30^{+0.06}_{-0.11}$ & $8.20^{+0.78}_{-0.16}$ & $68^{+7}_{-45}$ & $11\pm7$\\
4 & $-$ & $-$ & $73^{+11}_{-9}$ & $25\pm15$ \\
5 & $1.46^{+0.07}_{-0.17}$ & $8.59^{+0.35}_{-0.08}$ & $73^{+16}_{-10}$ & $34\pm7$ \\
6 & $1.50^{+0.18}_{-0.08}$ & $8.90^{+0.35}_{-0.08}$ & $77^{+10}_{-7}$ & $34\pm9$ \\
7 & $1.35^{+0.11}_{-0.11}$ & $8.78^{+0.43}_{-1.59}$ & $74^{+15}_{-9}$ & $10\pm7$ \\
8 & $-$ & $-$ & $72^{+13}_{-9}$ & $29\pm8$ \\
9 & $-$ & $-$ & $72^{+14}_{-9}$ & $25\pm6$ \\
10 & $1.09^{+0.04}_{-0.10}$ & $8.20^{+0.50}_{-0.08}$ & $65^{+23}_{-25}$ & $41\pm7$ \\
12 & $1.05^{+0.05}_{-0.14}$ & $8.90^{+0.27}_{-0.08}$ & $72^{+8}_{-10}$ & $87\pm9$ \\
\hline
\end{tabular}
\label{tab2}
\end{table}

Using the observed $P_l$ and the fitted emerging brightness
temperature we were then able to determine the angle between the
magnetic field and the maser propagation direction $\theta$. The
quoted errors on $\theta$ are the most compact 68 percent probability
interval, determined by analysing the full probability distribution
function. For those maser features where no radiative transfer fit was
possible, including the features without detected linear polarization,
we have used the range of brightness temperatures as suggested by the
fitted maser features. The inclination angles $\theta$ as well as the
angle on the plane of the sky $\phi_B$ for the maser features are
given in Table~\ref{tab2}. The errors in $\phi_B$ are determined from the
polarization angle $\chi$, with an added uncertainty of $5^\circ$ as described in \S~\ref{method}. As seen in Table~\ref{tab2} and shown in
Fig.~\ref{Fig:thetafit}, $\theta$ is constrained to
$\theta>\theta_{\rm crit}$ at better than $1\sigma$ level for the all but
three of the maser features. Of those three, the best fit $\theta$ for
features 3 and 10 are still larger than $\theta_{\rm crit}$. For the
majority of the maser features the magnetic field is thus likely
perpendicular to the measured polarization vectors. Only for feature
1, the best fit $\theta$ is less than $\theta_{\rm crit}$, meaning that the
magnetic field is more likely parallel to $\chi$.

\section{Discussion}
\label{disc}

\subsection{Methanol masers and the discs of Cepheus A HW2}

As noted in \S~\ref{dist}, the 6.7-GHz methanol masers lie in an
elliptical distribution around the molecular and dust discs of Cepheus
A HW2. The masers are found at $650$~AU, outside the molecular disc
with a radius of $580$~AU. The maser distribution also appears
$\sim20^\circ$ tilted with respect to the dust and molecular discs and
shows no sign of rotation. This indicates that the masers are not part
of the molecular disc itself. The temperature needed to maintain the
masing conditions, $T_k\sim200$~K \citep{Cragg05}, is slightly less
than the $T=250\pm30$~K found in the outer regions of the molecular
disc \citep{Jimenez09}. Also, the hydrogen number density in the maser
region, $n_{H_2}\sim10^9$~cm$^{-3}$ \citep{Cragg05} is approximately
three orders of magnitude less than that in the disc
\citep{Jimenez09}. Finally, the masers exhibit signs of infall and
their velocity range covers only $\sim3$~\kms, which indicates that it
is unlikely that they are part of the Cepheus~A~HW2 outflow. We thus
suggest that the maser, while not associated with the disc directly,
probe material that is being accreted onto the disc from the
surrounding medium.

\begin{figure}
\includegraphics[width=8.0cm]{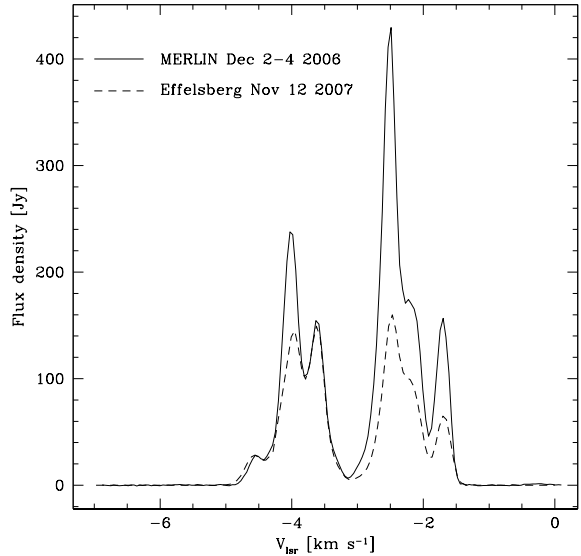}
\caption[spec]{Total intensity spectra of the MERLIN observations (solid line) and the Effelsberg observations (dashed line) that were used to determine the magnetic field strength in the 6.7-GHz methanol maser region \citep{Vlemmings08a}  }
\label{Fig:spec}
\end{figure}

\subsection{The magnetic field strength}
Dynamic range limits in our MERLIN observations did not enable us to
directly determine the magnetic field strength on the individual maser
features around Cepheus A~HW2 and we obtained a $3\sigma$ limit of
$\sim20$~mG for the line-of-sight magnetic field. However, the
magnetic field strength in the methanol maser region of Cepheus A was
measured using the Effelsberg 100-m telescope
\citep{Vlemmings08a}. Through Zeeman-splitting observations, the
line of sight field strength was determined to be $B_{\rm ||} = B
\cos(\theta) = 8.1\pm0.2$~mG \citep{Vlemmings08a}, which is mostly
constant across the full maser spectrum. As this value is a flux averaged
magnetic field determined with a single dish telescope, we need to
determine the amount of flux recovered and the number of maser
features detected in our higher resolution MERLIN observations, to
evaluate if the field measurement can be taken to be representative of
the observed maser features. In Fig.~\ref{Fig:spec} we show the
comparison between the MERLIN and Effelsberg spectra. It is clear that
the shape of the maser spectrum has changed significantly between the
two epochs, though all individual maser features found with MERLIN can
be identified in the Effelsberg spectrum. The change of the relative
flux of the individual maser features was noted in
\citet{Sugiyama08b}, who monitored Cepheus A with the Yamaguchi 32-m
telescope. Comparing our spectra with their figure 1, shows that the
MERLIN spectrum is remarkably similar to what was observed at August
12 2007, with no more than 10 percent of the flux resolved out at high
angular resolution. The MERLIN observations recover almost all of
the single dish maser flux, and all features in the Effelsberg
observations can be identified at high angular resolution. We can thus
conclude that the Effelsberg magnetic field measurement can be taken
as representative for true line-of-sight magnetic field in the maser
region. 

As discussed in \citet{Vlemmings08a}, there remains an estimated $25$
percent uncertainty on the magnetic field strength due to the
unknown accuracy of the Zeeman splitting coefficient of the 6.7-GHz
methanol maser transition. In addition to the measurement error of
$0.2$~mG on the magnetic field determination we thus add an additional
$2$~mG uncertainty to the magnetic field strength when we determine the
effect of the magnetic field around Cepheus A~HW2.

\subsection{Magnetic field morphology}
\begin{figure*}
\includegraphics[width=16.0cm]{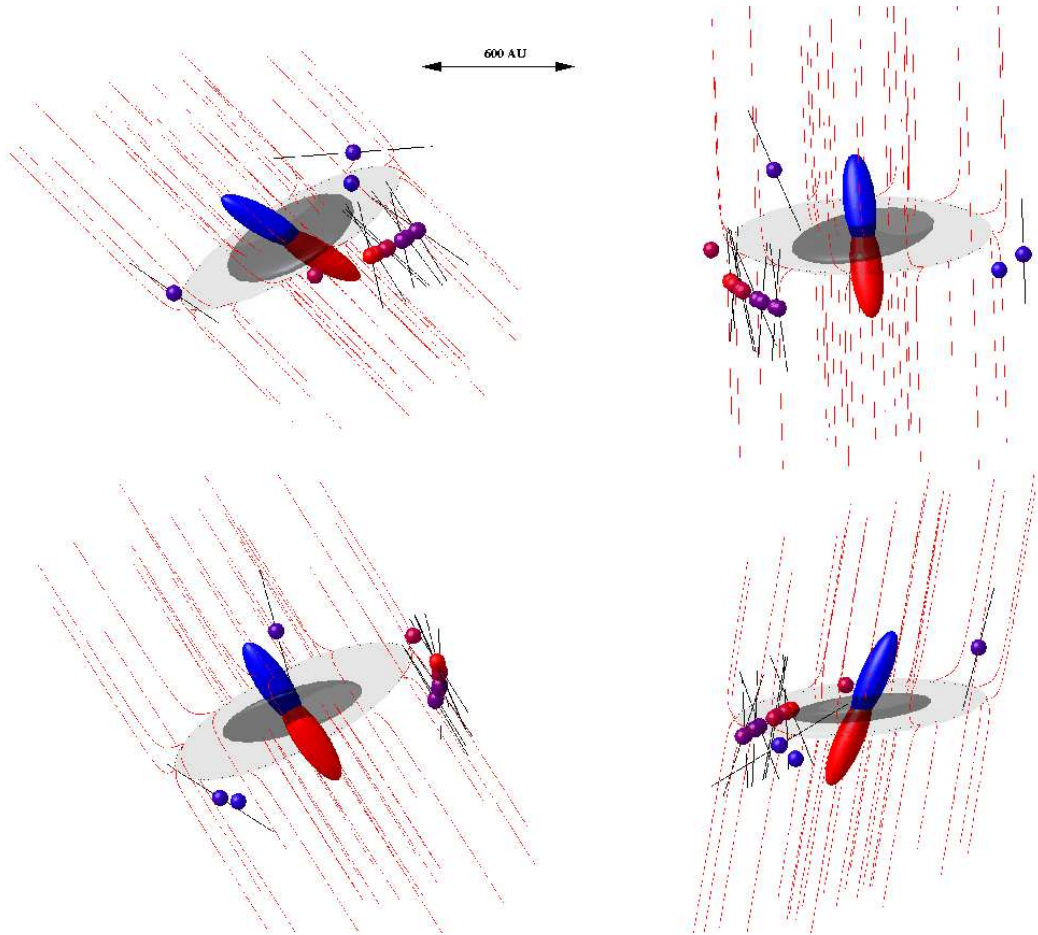}
\caption[model]{
The 3-dimensional magnetic field structure around the massive
protostar Cepheus A HW2. The top left panel corresponds to a viewing
angle close to the observations while the three further panels have
different viewing angles. Spheres indicate the masers, with the black
vectors indicating the true magnetic field direction. The maser
features are color-coded according to their velocity. The molecular
disc \citep{Patel05} is indicated by the dashed line and light grey structure
while the dust disc \citep{Jimenez07, Patel05} corresponds to the dark grey ellipse. The
blue- and red-shifted lobes of the collimated radio outflow \citep{Curiel06} are
also shown. The red lines indicate the proposed magnetic field
morphology.}
\label{Fig:cepa2}
\end{figure*}

The full 3-dimensional magnetic field orientation was inferred using
maser radiative transfer models described in \S~\ref{method} and is shown in Fig.~\ref{Fig:cepa2}. With the
exception of one of the maser features that originates from the
mid-plane of the molecular disc behind the outflow, the inferred
magnetic field in the plane of the sky is perpendicular to the
molecular and dust discs with a rms weighted average angle
$\phi_B=26\pm12^\circ$. Moreover, the error weighted average angle
between the magnetic field and the line of sight
$\theta=73\pm5^\circ$ is consistent with the inclination of the
molecular disc and the outflow. The overall magnetic field orientation
angle corresponds closely to the magnetic field direction observed in
the encompassing dust envelope \citep{Curran07}. The discrepant
polarization angle found on the mid-plane maser feature is either due
to the twisting of magnetic field lines towards the disc, or due to
Faraday rotation caused by the ionized outflow. While the methanol
masers at 6.7-GHz are not significantly affected by interstellar
Faraday rotation, an electron density of $\sim$200 cm$^{-3}$ along a
short 100 AU path through the ionized outflow with a few mG magnetic
field can cause the observed polarization angle of the masers behind
the outflow to rotate by more than 10$^\circ$.

\subsection{The role of the magnetic field}

Since we have obtained the angle between the magnetic field and the
line of sight, we can directly determine the magnetic field strength
around the protostellar disc. For $B_{\rm ||}=8.1\pm0.2\pm2$~mG and
$\theta=73\pm5^\circ$, the absolute field strength
$|B|=23^{+9}_{-7}$~mG. From this, we can calculate a number of
parameters to asses the role of the magnetic field during the massive
protostellar collapse. The Alfv{\'e}nic mach number $m_a =
\sigma\sqrt{3} / V_A$, where $\sigma = \Delta V/\sqrt{8\ln{2}}$ and
$\Delta V$ is the turbulent line-width. The fits to the radiative
transfer models indicate that the intrinsic maser line-width is much
larger than the thermal line-width expected in the maser region with a
typical kinetic temperature $T_k\sim200$~K. Assuming the intrinsic
line-width to be broadened due by turbulent motions, we find that
$\Delta V\sim1.1$~\kms. In the methanol maser region with a hydrogen
number density $n_{\rm H_2} \sim 10^9$~cm$^{-3}$, the Alfv{\'e}n
velocity $V_A = |B|/\sqrt{4\pi\rho}=2.05$~\kms. The measured infall
motions, at $1.7$~\kms, are thus slightly sub-Alfv{\'e}nic and $m_a =
0.4$. At 200~K the sonic mach number $m_s = 1.1$, and consequently,
the ratio between thermal and magnetic energy $\beta = 2 (m_a/m_s)^2 =
0.27$. The magnetic field dominates the energies in the high density
protostellar environment probed by the masers.

We can also calculate the mass to magnetic flux ratio $M/\Phi$
compared to the critical value of this ratio $ \lambda =
(M/\Phi)/(M/\Phi)_{\rm crit}$. The critical value $(M/\Phi)_{\rm
  crit}=c_\Phi/\sqrt{G}\approx0.12/\sqrt{G}$ \citep{Mouschovias76,
  Tomisaka88} is taken from numerical models for the collapse of a
spherical molecular cloud into a highly flattened structure and
defines the critical mass that can be supported against collapse by
the magnetic field. When $\lambda<1$, the magnetic field prevents
collapse, but when $\lambda$ becomes larger than 1 gravity overwhelms
the magnetic field. In terms of hydrogen column density, $\lambda$ can
be determined from $\lambda=7.6\times10^{-24} N(H_2)/|B|$, with the
magnetic field in mG. Taking the maser region to be $\sim$300 AU
thick, as determined by our observations, the hydrogen column density
along the magnetic field lines is $N(H_2)\approx
5\times10^{24}$~cm$^{-2}$. This yields $\lambda=1.7^{+0.7}_{-0.5}$, indicating
that the region is slightly super-critical, a condition needed for the
collapse to proceed. Similar values for $\lambda$ were found using
large-scale Zeeman-splitting observations of an ensemble of molecular
clouds \citep{Crutcher99}. However, as those observations lacked the
3-dimensional magnetic field information, the results were uncertain
by of order a factor two, making it impossible to distinguish between
sub- and super-critical cloud cores.

Finally, comparing the methanol maser magnetic field strength with
other field strength measurements in the high-density region around
HW2 ($n_{\rm H_2}>10^5$cm$^{-3}$), we find that the field scales
approximately as $B\propto n_{\rm H_2}^{0.5}$ \citep[Fig. 3 from][]{Vlemmings08a}. This relation could be the
consequence of ambipolar diffusion, but it also naturally occurs for
the collapse of a spherical cloud with frozen in field lines, when the
preferred infall direction is along the magnetic field. Such a
collapse forms a flattened disc-like structure similar as observed
around HW2. We thus suggest that the magnetic field plays a crucial
role in regulating the final stages of the formation of the massive
protostar Cepheus A HW2, as is commonly expected during low-mass
star-formation.

\subsection{Future perspectives}
\subsubsection{Cepheus A as a prototype methanol maser source?}
Due to their distance and complexity, the relation between the
methanol masers and the protostellar environment for most other maser
sources is still unknown. However, Zeeman-splitting observations have
shown typical magnetic field strengths along the line of sight of
$\sim$12 mG in over 70~percent of the observed methanol maser sources
\citep{Vlemmings08a}. Strong and dynamically important magnetic fields
are thus present in most massive star-forming region traced by
methanol masers. Furthermore, a recent survey has revealed that
$\sim$30~percent of the methanol maser sources display an elliptical
structure \citep{Bartkiewicz09}. This suggests that the methanol
masers of a large fraction of star forming regions are located in an interface
between a disc or torus and the infalling material, similar to the
inferred origin of the masers around Cepheus A. Our picture of
magnetic field regulated infall and accretion towards Cepheus A HW2
thus potentially describes a large number of massive star-forming
regions.

\subsubsection{Further e-MERLIN observations}
With the e-MERLIN upgrade \citep{Garrington04}, it will soon be
possible to map multiple maser transitions simultaneously with the
radio-continuum. By observing in full polarization mode, this will
allow for a direct comparison between the magnetic field structure
determined from the masers and the continuum morphology. Single track
observations will allow for mapping of massive star forming regions at
different evolutionary stages down to a continuum sensitivity
approaching $3$~$\mu$\jyb. Simultaneously, the sensitivity in a narrow
6.7-GHz maser frequency band will then be sufficient to detect linear
polarization for $>75$~percent of the star forming regions with
methanol masers when assuming the flux analysis of
\citet{vdWalt05}. When analyzed in a similar way to the Cepheus~A~HW2
observations presented here, such observations have the potential to
significantly further our understanding of the magnetic field around
massive protostars.

\section{Conclusion}

We have demonstrated the power of maser polarization observation, and
particularly those using methanol masers, in deducing the full
3-dimensional strength and structure of the magnetic field around
massive protostars. The detection of a coherent magnetic field
direction in maser features that individually have a size of a few AU,
suggests that the masers do not probe isolated pockets of a shock
compressed magnetic field. In that case, the polarization direction
would be determined by the direction of compression for each maser
individually and observing coherent polarization angles would be
unlikely. A similar conclusion could be drawn from recent observations
of W75N at high angular resolution with the EVN that reveal a uniform
magnetic field direction in individual methanol maser features spread
over $\sim2000$~AU \citep{Surcis09}. The detection of a large scale
magnetic field is further confirmed by the good agreement of the field
direction with that determined using dust polarization observations of
the entire Cepheus A region. The 3-dimensional magnetic field geometry
around Cepheus A HW2 supports the theories in which magnetic fields
regulate the infall and outflow close to massive protostars in a
similar way as during low-mass star-formation, even if the high-mass
star-forming regions are considerably more complex. The strong
magnetic field that appears to be threading the disc will play a
crucial role in maintaining the high accretion rate needed during
massive star-formation and is potentially essential in maintaining
disc stability and creating the conditions to allow for planet
formation \citep{Johansen05, Johansen08, Wardle07}.

\section*{Acknowledgments}

W.V. acknowledges support by the Deutsche Forschungsgemeinschaft (DFG)
through the Emmy Noether Research grant VL 61/3-1. G.S. is a member of
the International Max Planck Research School (IMPRS) for Astronomy and
Astrophysics at the Universities of Bonn and Cologne. K.T. was
supported by the EU Framework 6 Marie Curie Early Stage Training
programme under contract number MEST-CT-2005-19669 "ESTRELA" .

\bibliographystyle{mn2e}

%
%\bibliography{wvrefs}

\label{lastpage}

\end{document}